\input epsf
\documentstyle[twocolumn,twoside]{PAC95}

\def\BibTeX{{\rm B\kern-.05em{\sc i\kern-.025em b}\kern-.08em
    T\kern-.1667em\lower.7ex\hbox{E}\kern-.125emX}}
\def \lsa {\rlap {\lower 3.5 pt \hbox {$\mathchar \sim$}} \raise 1
pt \hbox {$<$}}
\def \rsa {\rlap {\lower 3.5 pt \hbox {$\mathchar \sim$}} \raise 1
pt \hbox {$>$}}

\begin{document}

\title{APPLICATIONS OF TIME DOMAIN SIMULATION TO COUPLER DESIGN FOR
PERIODIC STRUCTURES\thanks{Work supported by the Department of Energy,
contract DE-AC03-76SF00515, grants DE-FG03-93ER40759 and DE-FG03-93ER40793.}}

\author{
N.~M.~Kroll$^{1,2}$, C.-K.~Ng$^1$ and D.~C.~Vier$^2$\\
$^1$ Stanford Linear Accelerator Center, Stanford University, Stanford,
CA 94309, USA\\
$^2$ University of California, San Diego, La Jolla, 
CA 92093, USA\\
}

\maketitle

\thispagestyle{empty}\pagestyle{empty}

\begin{abstract} 
We present numerical procedures for analyzing the properties of
periodic structures and associated couplers based upon time domain
simulation.  Simple post processing procedures are given for determining
Brillouin diagrams and complex field distributions of the traveling wave
solutions, and the reflection coefficient of the traveling waves by the
input and output.  The availability of the reflection coefficient
information facilitates a systematic and efficient procedure for matching
the input and output.  The method has been extensively applied to coupler
design for a wide variety of structures and to a study directed towards
elimination of the surface field enhancement commonly experienced in
coupler cells.

\hskip 0.25 truein

\end{abstract}

\vspace{-0.75cm}
\section{Introduction}

\vspace{-0.1cm}

     Numerical simulation procedures for designing waveguide couplers to
accelerator structures are described in [1] and an example of its
application to the design of the input coupler for the NLC linac is given
in [2].  A coupler cavity is designed with the intent of providing a matched
connection between a waveguide and a uniform accelerator structure with
dimensions corresponding to those of the cell adjacent to the coupler
cavity.  A symmetric structure consisting of two coupler cavities (with
associated waveguides) connected by a short section of accelerator
structure (typically two cells worth) is modeled and subjected to a (let us
assume single frequency) time domain simulation.  The entire assembly is
treated as a single structure with two wave guide ports.  The coupler cell
dimensions are adjusted until an apparent match is achieved, that is, until
no reflection is experienced at the ports (the external matching
condition).  To eliminate the possibility that the match arises from a
fortuitous cancellation between forward and backward waves within the
accelerator structure, both the amplitude and phase of the accelerating
field on the beam axis are observed and required to have the periodicity
and phase advance properties appropriate to a pure traveling wave (the
internal matching condition).  As a check one may add a cell to the
accelerator structure and see whether all these conditions are still
satisfied.  It is often the case that accelerator structures are slowly
varying rather than uniform, in which case the input and output couplers
are matched separately.

     In practice the procedure (we refer to it as the standard procedure)
has been quite time consuming, involving trial and error rather than a
systematic procedure to simultaneously satisfy both the internal and
external matching conditions.  Another limitation arises from the fact that
the method assumes that evanescent bands can be neglected but provides no
procedure for demonstrating their absence.

     In the next section we describe a new simulation procedure which has
been found to be much more efficient, and also which provides information
about the presence of evanescent bands.  The basic elements of the method
were briefly described in [3] in connection with the design of a coupler
for the zipper structure.  Because it has since replaced the old method for
all of our coupler design work, a more complete presentation together with
examples will be presented in the following sections.

\section{The New Simulation Procedure}

     As in the case of the old standard procedure one applies a single
frequency time domain simulation by driving the input port of a two port
structure consisting of an input cavity, an $N$ cell periodic structure with
period $P$, and an output cavity.  Instead, however, of focusing attention
on the S parameters of the structure as a whole, we direct our attention to
the simulated accelerating field $E_z(z,t)$ evaluated along the beam axis.
We assume a steady state has been reached, so that the subsequent time
dependence can be expressed in terms of the complex $E_z(z)$ ($E_c(z)$
henceforth), obtained in the standard way by combining the simulated real
fields at two times separated by a quarter period.  Then from Floquet's
theorem (neglecting evanescent bands, losses, and an irrelevant overall
phase factor)

\begin{equation}
       E_c(z) = E(z)[exp(-j\phi(z)) + R exp(j\phi(z))].
\end{equation}

\noindent
Here $E(z)$ is a real positive amplitude function with period $P$, and $\phi(z)$
is a real phase function, periodic except for a cell to cell phase advance
$\psi$.  Thus

\begin{equation}
       E(z \pm P) = E(z),  \; {\rm and} \;   
       \phi(z \pm P) = \phi(z) \pm \psi .    
\end{equation}

\noindent
$R$ is a $z$ independent complex reflection coefficient.  Note that one
is free to shift $\phi$ by an arbitrary constant with a compensating phase
shift in $R$, since the overall phase of $E_c$ is irrelevant. This freedom
corresponds to the choice of reference plane through some point $z_0$ where
we take $\phi = 0$.

We now consider the quantities
\begin{equation}
      \Sigma(z) = F^+(z) + F^-(z),  \; {\rm and} \; \Delta(z) = F^+(z) - F^-(z), 
\end{equation}

\noindent
where  \begin{equation}
F^\pm(z) = E_c(z \pm P)/E_c(z).                                
\end{equation}

\noindent
Elementary algebraic manipulation leads to the relations:

\begin{equation}
       2Cos(\psi) = \Sigma(z),
\end{equation}
\begin{equation}
       R exp(2j\phi) = [2Sin(\psi)-j\Delta(z)]/[2Sin(\psi)+j\Delta(z)].
\end{equation}

\noindent
We note that while the RHS of (5) is formed of $z$ dependent complex
quantities, it nevertheless turns out to be real and $z$ independent.
Similarly the absolute value of the RHS of (6) is also $z$ independent.  Both
these results should hold for all ``allowed'' $z$ values, i.e., values such 
that the three points $z+P$, $z$, and $z-P$ all lie within the periodic portion
of the structure simulated, and together they constitute a powerful constraint 
on the validity of the Floquet representation Eq.~(1).  Their failure beyond
small numerical fluctuations or small deviation from steady state is
evidence for the presence of evanescent bands.  An example will be presented
in the section on the Zipper structure.

     It is noteworthy that these relations allow one to determine all the
properties of the traveling wave solutions, including the functions $\phi(z)$
and $E(z)$ from a simulation which contains an arbitrary mixture of forward
and backward waves.  Of particular importance is the fact that it gives the
magnitude and phase of the reflection coefficient.  In contrast to the old
standard method, there is here only one matching condition to be satisfied,
namely $|R| = 0$.  Typically match is achieved by varying two parameters in the
coupler design.  Once one has determined how the real and imaginary parts
of $R$ vary with the parameters, one can choose linear combinations of
changes which accelerate the process of converging to the origin in the
complex $R$ plane [4].  Because the phase of $R$ does depend upon the position 
of the reference point relative to the couplers, one naturally keeps it fixed
while carrying out this process.  Note that it is the output cavity that is
matched by this procedure.  While not necessary, it is usually convenient
to construct a symmetric mesh.  The input and output cavities are then the
same, and the structure as a whole is matched when $R$ vanishes.

\section{Applications and Examples}

\noindent
{\it (a)  The NLC four port output coupler}

     As an example of the principal features of the new method we use the
new NLC four port output coupler cavity [5].  The purpose of the four port
design was to provide damping for those dipole modes that reach the end of
the structure while also providing an output for the accelerating mode.
These dipole modes are typically those which had been poorly damped because
of decoupling of the last cells from the manifolds.  The four port symmetry
provides damping for both dipole mode polarizations and has the added
advantage of eliminating quadrupole distortion of the coupler fields.

     The design simulation was carried out with a three cell periodic
structure, and results are illustrated in Fig.~(1).  Two cases are shown,
one matched, the other not. The reflection coefficients $|R|$ as computed
from Eq.~(5) for the two cases are shown as functions of $z$. The 
allowed z values are those lying within the central cell, and one sees that 
for both cases $|R|$ is constant over that range.  The real part of 
$Cos(\psi)$ is also plotted as a function of $z$. One sees that the two 
values are indeed constant over the allowed range, but contrary to expectations 
they differ somewhat from each other and from the expected value of one half.
This is due to the fact that a different and coarser mesh than that used to 
determine the phase advance parameter was used for the time domain 
simulations. The two cases differ from one another because the parameter 
variations in the coupler associated with the matching procedure induce 
small but global changes in the meshing. It has been confirmed in a number 
of cases that there is good agreement between the phase frequency relation 
determined from single cell periodic boundary condition frequency domain
calculations and that determined from the time domain method described here 
so long as the same mesh is used for both simulations. 

\begin{figure}[h]
  \begin{picture}(100,200)(-10,15)
         { \epsfxsize=7.5cm
           \epsfbox{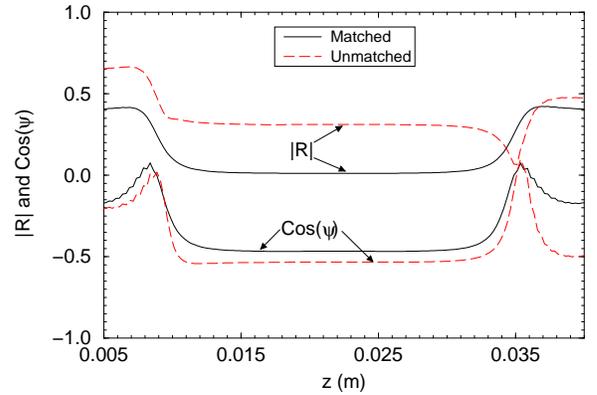}
         }
  \end{picture}

\vspace{-2.25cm}
\caption{$|R|$ and $Cos(\psi)$ along the axis of the NLC four-port 
output coupler.}
\label{nlc4out}
\end{figure}
\vspace{-0.35cm}

\noindent
{\it (b)  A Photonic Band Gap (PBG) structure}

\begin{figure}[h]
  \begin{picture}(100,200)(-5,-65)
         { \epsfxsize=8.5cm
           \epsfbox{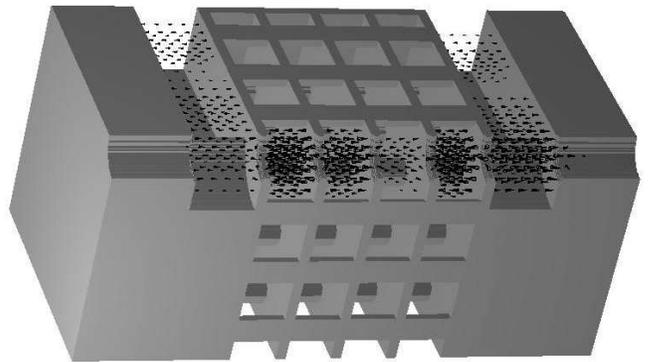}
         }
  \end{picture}

\vspace{-2.25cm}
\caption{A snapshot of electric field in the PBG structure.}
\label{transition}
\end{figure}
\vspace{-0.35cm}

    A coupler cell very similar to those of the SLAC structures has been
designed for a PBG structure, that is, a cylindrical cell with a pair of 
symmetrically placed
waveguide ports, a conventional beam pipe, and conventional beam iris
coupling to the periodic PBG structure.  The PBG cell structure [6] is a
seven by seven square array of metallic posts aligned in the beam direction
and terminated by metallic end plates, the cell cavity being formed by
removing the central post.  A circular aperture in the end plates,
identical to that between the coupler cell and the adjacent PBG cell,
provides cell to cell coupling and a path for the beam.  A perspective
representation of the four cell quarter structure used for the simulations 
is shown in Fig.~(2).  Also shown is the simulated electric field 
distribution, scaled logarithmically to enhance the visibility of weak field 
strengths. The figure illustrates the effectiveness of the PBG structure in 
confining the acceleration fields to the interior of the structure. 
The matching procedure worked well, and, as in the four port coupler
above, there was no evidence for evanescent band contamination.  Fabrication
of an experimental model with 5 coupled PBG cells and complete with couplers 
is in progress at SLAC.

\vspace{0.1cm}
\noindent
{\it (c)  The Zipper structure}

     The zipper is a planar accelerator structure described in [3].  A 25
(counting the coupler cavities) cell W band model has been built, cold
tested, and subjected to bead pull measurements as reported in [7].  The
design was governed by a decision to avoid bonded joints involving tiny
structure elements such as the vanes which serve as cell boundaries and
also form the beam iris.  The coupler cell is a quarter wave transformer
terminating in WR10 waveguide.  

     Early attempts at matching the coupler using the old standard method
failed, and it was this failure which led to the development reported
here.  Matching using this method was accomplished by making use of a time
domain simulation of a structure with 22 periodic cells.  Fig.~3 shows
the resultant $Re \, Cos(\psi)$, $Im \, Cos(\psi)$,  and $|R|$ plots as 
computed from Eqs.~(5) 
and (6). One sees large deviations from the expected $z$ independent behavior 
as one moves away from the center of the structure.  This effect indicates a 
clear violation of Eq.~(1).  From the fact that the violation fades away as one
moves away from the couplers indicates that the effect is due to the 
couplers generating an evanescent band, the nearby monopole band pointed
out in [3].  This example demonstrates how the method indicates the
presence of evanescent band interference, and also how one can carry out
the matching procedure even when it is present.

\begin{figure}[h]
  \begin{picture}(100,200)(-5,-62.5)
         { \epsfxsize=8.cm
           \epsfbox{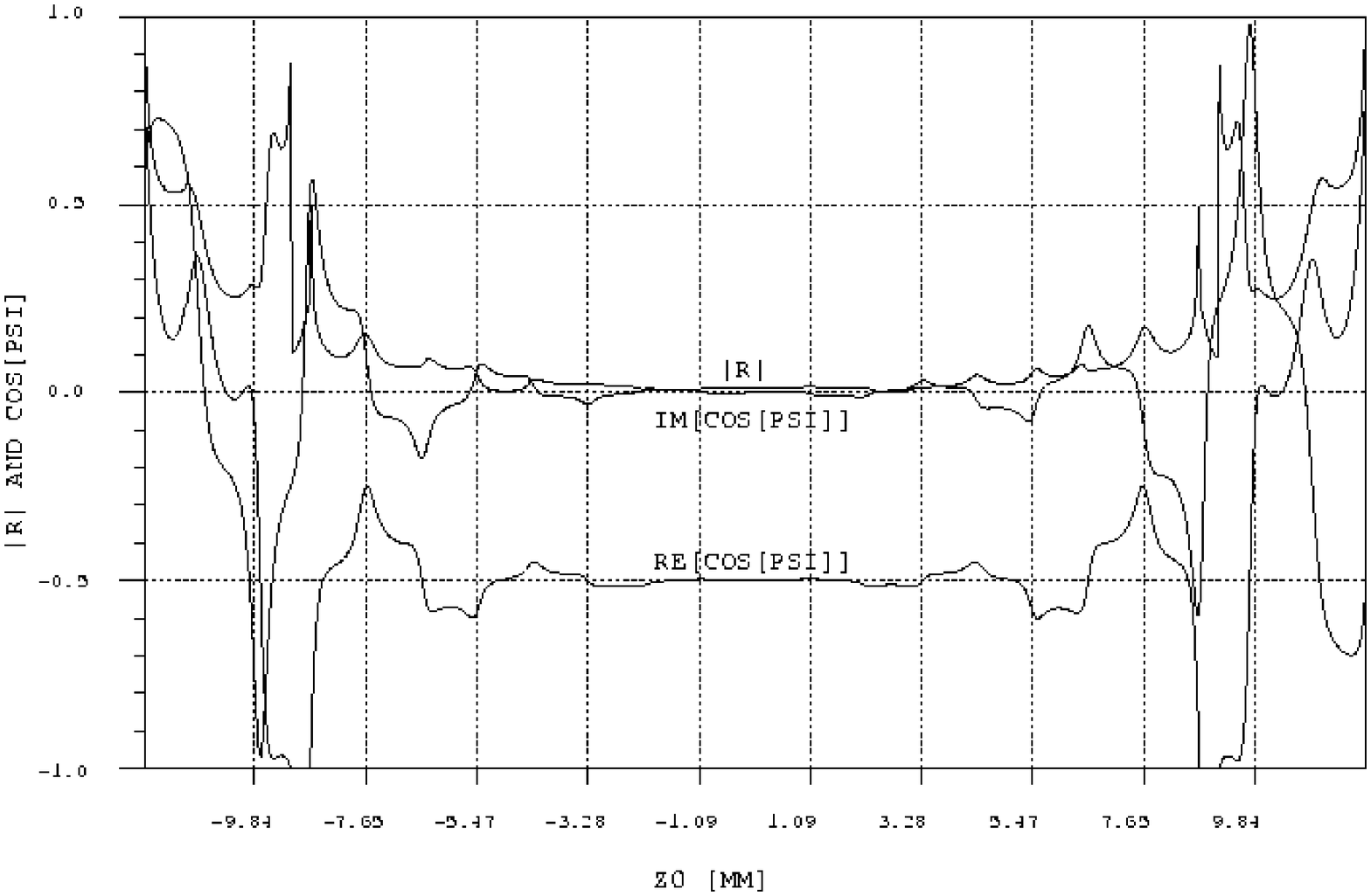}
         }
  \end{picture}

\vspace{-2.4cm}
\caption{$|R|$ and $Cos(\psi)$ along the axis of the zipper structure.}
\label{transition}
\end{figure}
\vspace{-0.1cm}

\section{The Coupler Field Enhancement Problem}

     Electrical discharge damage has been commonly observed in the coupler
cells of accelerator structures and has been attributed to the field
enhancement noted in simulations.  We have taken advantage of our enhanced
matching capability to initiate a study of this long standing problem.
Exploration of the situation for the NLC coupler [2] showed that the
largest enhancement occurred on the coupler side of the aperture of the
iris separating the coupler from the adjacent cell with azimuthal maxima
opposite the coupler waveguides and azimuthal minima 90 degrees away.
This observation was consistent with the pattern of discharge damage [9].  It
is pointed out in [8] that the azimuthal variation is due to the quadrupole
component introduced by the coupler waveguides and that the enhancement can
be reduced by introducing a racetrack like modification of the coupler cell
shape designed to eliminate it.  This effect and its cure have been
confirmed in our own studies of the NLC coupler.  Two other modifications
have also been explored.  The simplest and most effective was simply to
reduce the radius of the cell adjacent to the coupler.  The effect for a 2$\%$
reduction is illustrated in Fig.~4 where it is seen that the field on the
coupler cell iris is significantly less than that on the interior coupling
irises.  An undesirable consequence is a 10 degree phase advance
deficiency in the modified cell.  An even larger field reduction would be
obtained by removing the quadrupole enhancement.  We attribute the
reduction to an increase in group velocity.  The other modification
consisted of enlarging the coupler iris combined with an increase in the
adjacent cell radius chosen so as to preserve the cell phase advance, but
the exploration of this effect is incomplete.  Experimental investigation
to determine whether such changes actually do reduce electrical discharge
damage in the coupler is clearly needed.

\begin{figure}[h]
  \begin{picture}(100,200)(-10,+15)
         { \epsfxsize=7.5cm
           \epsfbox{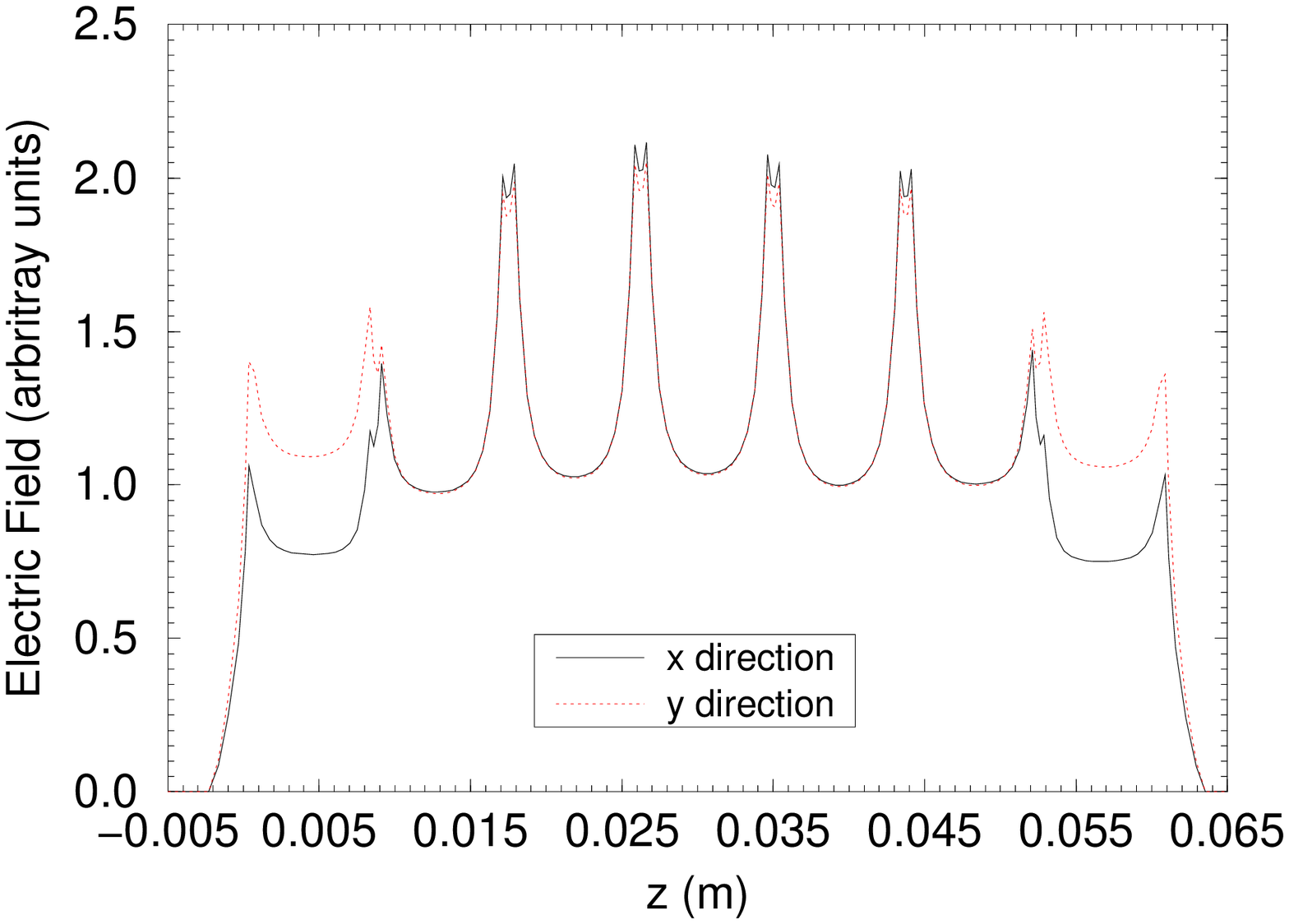}
         }
  \end{picture}

\vspace{-2.2cm}
\caption{$z$ variation of electric field magnitude at radial positions of
the beam irises.}
\label{transition}
\end{figure}

\end{document}